\documentclass[12pt,a4paper]{article}
\usepackage{amssymb}
\usepackage{epsfig}
\begin{document}
\null
\begin{flushright}
\begin{tabular}{r}
DFTT 61/98\\
hep-ph/9810272\\
October 6, 1998
\end{tabular}
\end{flushright}
\vspace{0.5cm}
\begin{center}
\Large \bfseries
Is bi-maximal mixing compatible with
the large angle MSW solution of the solar neutrino problem?
\\[0.5cm]
\large \mdseries
C. Giunti
\\[0.25cm]
\itshape
\normalsize
INFN, Sezione di Torino, 
and
Dipartimento di Fisica Teorica,
\\
Universit\`a di Torino,
Via P. Giuria 1, I--10125 Torino, Italy
\\
\vspace{0.5cm}
\upshape
\large
Abstract
\\[0.25cm]
\normalsize
\begin{minipage}[t]{0.9\textwidth}
It is shown that the large angle MSW solution of the solar neutrino problem
with a bi-maximal neutrino mixing matrix
implies an energy-independent suppression of the solar $\nu_e$ flux.
The present solar neutrino data exclude this solution of the
solar neutrino problem at 99.6\% CL.
\end{minipage}
\end{center}

\newpage

The possibility that the neutrino mixing matrix $U$ has the
bi-maximal mixing form
\begin{equation}
U
=
\left(
\begin{array}{ccc}
\frac{1}{\sqrt{2}} & -\frac{1}{\sqrt{2}} & 0
\\
\frac{1}{2} &  \frac{1}{2} & \frac{1}{\sqrt{2}}
\\
\frac{1}{2} &  \frac{1}{2} & -\frac{1}{\sqrt{2}}
\end{array}
\right)
\label{001}
\end{equation}
has attracted a large attention \cite{bi-maximal}
after the presentation at the \textit{Neutrino '98}
conference of the Super-Kamiokande evidence
in favor of atmospheric neutrino oscillations
with large mixing \cite{SK-atm}.

Neutrino bi-maximal mixing is capable of explaining in a elegant way
the atmospheric neutrino anomaly
\cite{SK-atm,Kam-atm-94,IMB95,Soudan97},
through
$\nu_\mu\to\nu_\tau$
oscillations due to\footnote{$ \Delta{m}^2_{kj} \equiv m_k^2 - m_j^2 $
is the difference between the squared masses of the two massive neutrinos
$\nu_k$ and $\nu_j$.
In the bi-maximal mixing scenario there are three massive neutrinos,
$\nu_1$, $\nu_2$ and $\nu_3$.}
$ \Delta{m}^2_{31} \sim 10^{-3} \, \mathrm{eV}^2 $
and the solar neutrino problem (SNP)
\cite{Homestake98,Kam-sun-96,GALLEX96,SAGE96,SK-sun-nu98}
through
$\nu_e\to\nu_\mu,\nu_\tau$
oscillations in vacuum due to
$ \Delta{m}^2_{21} \sim 10^{-10} \, \mathrm{eV}^2 $
\cite{BG98}.

As noted in \cite{Mohapatra-Nussinov98b},
the results of the recent analysis of solar neutrino data
presented in \cite{BKS98}
seem to imply\footnote{I want to emphasize from the beginning that
I do not want to criticize at all the beautiful paper \cite{BKS98}.
I am only concerned with the interpretation of its results.}
that neutrino bi-maximal mixing
may be also compatible at 99\% CL
with the large mixing angle (LMA) MSW \cite{MSW}
solution of the SNP \cite{SNP}
(see Fig.~2 of \cite{BKS98}).

Here I would like to notice that this conclusion seems to be in contradiction
with the exclusion at 99.8\% CL
of an energy-independent suppression of the solar $\nu_e$ flux
presented in the same paper \cite{BKS98}
(see section IV.D).

The reason of this incompatibility is that
bi-maximal mixing with the
$ \Delta{m}^2_{21} \sim 10^{-5} - 10^{-4} \, \mathrm{eV}^2 $
corresponding to the LMA solution of the SNP
implies an energy-independent suppression by a factor 1/2
of the solar $\nu_e$ flux.

This can be seen following the simple reasoning presented in \cite{BGK96b}.
The mixing of the neutrino states in vacuum is given by
(see, for example, \cite{no-revs})
\begin{equation}
|\nu_\alpha\rangle
=
\sum_{k=1,2,3} U_{{\alpha}k}^* \, |\nu_k\rangle
\qquad
(\alpha=e,\mu,\tau)
\,,
\label{mixing}
\end{equation}
where the states
$|\nu_\alpha\rangle$ ($\alpha=e,\mu,\tau$)
describe neutrinos produced in weak interaction processes
and
the states
$|\nu_k\rangle$ ($k=1,2,3$)
describe neutrinos with definite masses $m_k$.

In the bi-maximal mixing scenario the numbering of the
massive neutrinos is the usual one,
\textit{i.e.} such that
$ m_1 \leq m_2 \leq m_3 $,
and
$ \Delta{m}^2_{31} \sim 10^{-3} \, \mathrm{eV}^2 $
for the solution of the atmospheric neutrino anomaly.
If
$ \Delta{m}^2_{21} \sim 10^{-5} - 10^{-4} \, \mathrm{eV}^2 $
for the LMA solution of the SNP,
we have
$ \Delta{m}^2_{32} \simeq \Delta{m}^2_{31} \sim 10^{-3} \, \mathrm{eV}^2 $.

Solar neutrinos have energy $ E \sim 1 \, \mathrm{MeV} $
and the ratio
$
\Delta{m}^2_{31} / E
\simeq
\Delta{m}^2_{32} / E
\sim
10^{-9} \, \mathrm{eV}
$
is much larger than the matter induced potential
$ V \lesssim 10^{-11} \, \mathrm{eV} $
in the interior of the sun.
Hence,
the evolution equation of the heaviest massive neutrino $\nu_3$
is decoupled from that of the two light neutrinos
$\nu_1$ and $\nu_2$
(see, for example, \cite{Kuo-Pantaleone89}).
Taking also in account that in the case of bi-maximal mixing $U_{e3}=0$,
one can see that an electron neutrino is created in the core of the sun as
a superposition of the two light mass eigenstates
$\nu_1$ and $\nu_2$
and,
whatever happens during his propagation in the interior of the sun,
its state when it emerges from the surface
of the sun is a linear combination of
$|\nu_1\rangle$
and
$|\nu_2\rangle$:
\begin{equation}
|\nu\rangle_S
=
\sum_{k=1,2} a_k \, |\nu_k\rangle
\,,
\label{002}
\end{equation}
with
\begin{equation}
|a_1|^2 + |a_2|^2 = 1
\,.
\label{003}
\end{equation}
Since the massive neutrino states $|\nu_k\rangle$
propagate as plane waves,
the state describing the neutrino detected on the Earth is
\begin{equation}
|\nu\rangle_E
=
\sum_{k=1,2} a_k \, e^{-i E_k L} \, |\nu_k\rangle
\,,
\label{004}
\end{equation}
where $L$ is the distance from the surface of the Sun to the detector
on the Earth.
The survival probability of solar electron neutrinos is then given by
$ P_{\nu_e\to\nu_e}^{\mathrm{sun}} = | \langle\nu_e|\nu\rangle_E |^2 $:
\begin{equation}
P_{\nu_e\to\nu_e}
=
\left| \sum_{k=1,2} a_k \, e^{-i E_k L} \, \langle\nu_e|\nu_k\rangle \right|^2
=
\left| \sum_{k=1,2} a_k \, e^{-i E_k L} \, U_{ek} \right|^2
\,.
\label{005}
\end{equation}
Taking now into account the explicit values
$ U_{e1} = 1 / \sqrt{2} $
and
$ U_{e2} = - 1 / \sqrt{2} $
in the case of bi-maximal mixing
and the fact that the neutrinos are extremely relativistic,
we have
\begin{equation}
P_{\nu_e\to\nu_e}
=
\frac{1}{2}
\left|
a_1 - a_2 \, \exp\left( - i \, \frac{ \Delta{m}^2_{21} L }{ 2 E } \right) 
\right|^2
\,.
\label{006}
\end{equation}
In the case of the LMA solution of the SNP
$ \Delta{m}^2_{21} \sim 10^{-5} - 10^{-4} \, \mathrm{eV}^2 $
and the survival probability (\ref{006}) oscillates with an oscillation length
$ 
4 \pi E / \Delta{m}^2_{21}
\sim
10^7 \, \mathrm{cm}
$
that is about one million times smaller than the Sun--Earth distance.
Hence,
the oscillations are not observable on the Earth because of averaging
over the energy spectrum
and only the average probability
\begin{equation}
\langle P_{\nu_e\to\nu_e} \rangle
=
\frac{1}{2} \, ( |a_1|^2 + |a_2|^2 )
=
\frac{1}{2}
\label{007}
\end{equation}
is observable.
We have obtained the announced result:
\emph{the LMA solution of the SNP in the bi-maximal mixing scenario
implies an energy-independent suppression of the solar $\nu_e$ flux
of a factor $1/2$}.

Therefore,
we have the apparent paradox that
an energy-independent suppression of the solar $\nu_e$ flux
seems to be allowed at 99\% CL
by Fig.~2 of Ref.~\cite{BKS98}
and is excluded at 99.8\% CL
in Section IV.D of the same paper.
Notice that the two conclusions are based on the same set of data
and the same theoretical calculation
of the neutrino flux produced by thermonuclear reactions in the core of the sun
\cite{BP98}.

The fact that the two cases refer to the same physical situation,
\textit{i.e.}
an energy-independent suppression of the solar $\nu_e$ flux,
is also shown by the $\chi^2$ calculated in the two cases.
The $\chi^2$ of the right border of the LMA region\footnote{If
$U_{e3}=0$,
we have
$\sin^22\vartheta=4|U_{e1}|^2|U_{e2}|^2$
(see \cite{BG98})
and
$\sin^22\vartheta=1$
corresponds to
$|U_{e1}|=|U_{e2}|=1/\sqrt{2}$,
as in the bi-maximal mixing matrix (\ref{001}).}
in Fig.~2 of Ref.~\cite{BKS98}
is $ 4.3 + 9.2 = 13.5 $,
whereas the $\chi^2$
calculated in Section IV.D of the same paper
for an energy-independent suppression of the solar $\nu_e$ flux
by a factor $0.48$
is 12.0.
Since this is the best fit
for an energy-independent suppression of the solar $\nu_e$ flux,
a value of
$\chi^2 = 13.5$
for a suppression factor $0.5$ looks plausible.

The solution of the apparent paradox explained above
lies in
\emph{a correct statistical interpretation of the
allowed LMA region in Fig.~2 of Ref.~\cite{BKS98}
and of the exclusion in Section IV.D of the same paper}.
The two cases have different statistical meanings.

The allowed regions in Fig.~2 of Ref.~\cite{BKS98}
are obtained under the assumption that
the neutrino masses and mixing parameters are not known.
In this case a general neutrino oscillation formula is used in the fit,
with the neutrino masses and mixing angles considered as free parameters.
The best fit in the LMA region happens to have a
$\chi^2_{\mathrm{min}}=4.3$,
which corresponds to a CL of 3.8\%
with 1 DOF.
Hence,
a LMA solution is allowed at 3.8\% CL
and one can draw a 99\% CL
region corresponding to the parameters that have
$\chi^2 \leq \chi^2_{\mathrm{min}} + 9.2 $.

The statistical analysis discussed
in Section IV.D of Ref.~\cite{BKS98}
assumes that the solar $\nu_e$ flux is suppressed by a constant factor
that is the free parameter to be determined by the fit.
It happens that the best fit has
$\chi^2_{\mathrm{min}}=12.0$,
which corresponds to a CL of 0.2\%
with 2 DOF.
Hence, the hypothesis is excluded at 99.8\% CL
and no allowed region of the free parameter can be drawn.

Since the two statistical analyses
start from different assumptions,
it is clear that they answer different questions
and their conclusions cannot be compared.
Moreover,
it is important to notice that the test of the maximal mixing scenario
with
$ \Delta{m}^2_{21} \sim 10^{-5} - 10^{-4} \, \mathrm{eV}^2 $
does not correspond to either of the two statistical analyses.
Indeed,
if this scenario is assumed,
we \textit{know}
that the solar $\nu_e$ flux is suppressed by an energy-independent
factor 0.5
and there is no parameter to fit.
Hence we test the hypothesis under consideration
on the basis of its $\chi^2$.
The $\chi^2\simeq13.5$
indicated by Fig.~2 of Ref.~\cite{BKS98}
implies a CL of 0.4\%
with 3 DOF.
Therefore,
the hypothesis is rejected at 99.6\% CL.

Notice that this exclusion is based only on the values of the elements
$U_{e1}$, $U_{e2}$ and $U_{e3}$
of the neutrino mixing matrix.
This means that also other types of neutrino mixing matrix,
as those discussed in \cite{other-sun-maximal},
are incompatible with the LMA solution of the SNP.

In conclusion,
I would like to emphasize that the allowed regions
of the neutrino oscillation parameters calculated in the usual way
(\textit{i.e.} as Fig.~2 of Ref.~\cite{BKS98})
cannot be used to test a definite model
(as the bi-maximal mixing model)
because they have been obtained under different
assumptions\footnote{They are useful if one wants to know
the allowed range of the mixing parameters for other purposes.}.
In order to obtain allowed regions
appropriate for model testing
one must use the procedure described in \cite{Krauss95,BG94-analysis},
\textit{i.e.}
one must consider each point of the parameter space as a model
and perform a goodness of fit testing with it.
I think that it would be very useful if
both types of allowed regions
will be presented in future papers.

\bigskip

\begin{flushleft}
\textbf{Acknowledgement}
\end{flushleft}

I would like to thank Z.Z. Xing for bringing my attention to the problem
under discussion.


\end{document}